\documentclass[letterpaper,twocolumn,10pt]{article}
\usepackage{usenix2019_v3}
\usepackage[utf8]{inputenc}
\usepackage{amsmath}
\usepackage{amsthm}
\usepackage{amssymb}
\usepackage{enumitem}
\usepackage{microtype}
\usepackage{graphicx}
\usepackage{subfigure}
\usepackage{booktabs} 
\usepackage{xspace}
\usepackage{bm}
\usepackage{bbm}
\usepackage{cite}
\usepackage{hyperref}
\usepackage{xcolor} 
\usepackage{algorithm}
\usepackage{algorithmic}
\usepackage{bold-extra}
\usepackage{makecell}

\newcommand{\vct}[1]{\bm{#1}}

\newcommand{\scheme}{\texttt{OmniLytics+}\xspace}
\newcommand{\contract}{\texttt{ModelTrade}\xspace}
\newcommand{\ethcontract}{\texttt{EthModelTrade}\xspace}
\newcommand{\ethe}{\texttt{Ethereum}\xspace}

\newcommand{\mr}{\texttt{MO}\xspace}
\newcommand{\dr}{\texttt{DO}\xspace}
\newcommand{\drs}{\texttt{DO}s\xspace}

\usepackage{tikz}

\usepackage{filecontents}

\begin{document}
\sloppy
\date{}

\title{\scheme: A Secure, Efficient, and Affordable Blockchain Data Market for Machine Learning through Off-Chain Processing}

\author{
{\rm Songze Li}\\
Southeast University
\and
{\rm Mingzhe Liu}\\
HKUST(GZ)
\and
{\rm Mengqi Chen}\\
HKUST(GZ)
} 

\maketitle

\begin{abstract}
The rapid development of large machine learning (ML) models requires a massive amount of training data, resulting in booming demands of data sharing and trading through data markets. Traditional centralized data markets suffer from low level of security, and emerging decentralized platforms are faced with efficiency and privacy challenges. In this paper, we propose \scheme, the first decentralized data market, built upon blockchain and smart contract technologies, to simultaneously achieve 1) data (resp., model) privacy for the data (resp. model) owner; 2) robustness against malicious data owners; 3) efficient data validation and aggregation. Specifically, adopting the zero-knowledge (ZK) rollup paradigm, \scheme proposes to secret share encrypted local gradients, computed from the encrypted global model, with a set of untrusted off-chain servers, who collaboratively generate a ZK proof on the validity of the gradient. In this way, the storage and processing overheads are securely offloaded from blockchain verifiers, significantly improving the privacy, efficiency, and affordability over existing rollup solutions. We implement the proposed \scheme data market as an Ethereum smart contract~\cite{Omni}. Extensive experiments demonstrate the effectiveness of \scheme in training large ML models in presence of malicious data owner, and the substantial advantages of \scheme in gas cost and execution time over baselines.
\end{abstract}

\section{Introduction}
Recent years have witnessed explosive developments of machine learning (ML) in various industry sectors like autonomous driving, content generation, and healthcare. As the ML models become increasingly large in size (e.g., the GPT-3 language model has 175 billion parameters~\cite{brown2020language}), obtaining high-quality models that satisfy business needs requires massive amount of training data, which makes data a valuable resource to fuel the AI industry. Under this background, data markets naturally emerge as an intermediate platform, for model owners to purchase intended data to facilitate their model training, and for data owners to trade the value of their private data for profit~\cite{liang2018survey}.

Conventional data markets are often implemented in a centralized fashion as a cloud application, on which data providers and consumers register themselves as users. To facilitate effective and satisfactory data trading, a suite of mechanisms are designed and implemented on the market to realize desirable functionalities like data pricing and matching between providers and consumers~\cite{krishnamachari2018i3,mivsura2016data,hayashi2020teeda,fernandez2020data}. While simple to implement, centralized designs are faced with severe security threats, such that the entire service is disrupted when several cloud servers or a datacenter is compromised. This security concern spurs the recent paradigm shift for data market design towards decentralized architectures, especially blockchains~\cite{ozyilmaz2018idmob,banerjee2018blockchain,ramachandran2018towards,koutsos2021agora,hynes2018demonstration}. A blockchain system (e.g., Bitcoin~\cite{nakamoto2008bitcoin} and Ethereum~
\cite{wood2014ethereum}) consists of a large number of decentralized nodes, who agree on and execute sequences of transactions through consensus algorithms and cryptographic tools, even when a subset of nodes are corrupted by adversary. Decentralized data market requires substantially more resources for attackers to take over, hence significantly enhancing the security of the market.

Blockchain-based data markets leverage smart contracts~\cite{ETH} to securely implement the trading rules. A naive implementation would have the data providers and the consumers include their data and requests respectively in blockchain transactions, which will be uploaded and processed by blockchain nodes (or verifiers). This is faced with two major challenges for building an ML data market: 1) efficiency and affordability: the size of the training data or ML model parameters is quite large, which would incurs a prohibitively high computational cost for the verifiers and monetary cost for data trading participants; 2) privacy: as the contents of the transactions are transparent to public, private data and proprietary model parameters may be leaked. To address the first challenge, many existing solutions choose to store the data onto a remote file system (e.g., IPFS~\cite{benet2014ipfs}), and record the commitment of the data on-chain (see, e.g.,~\cite{ramachandran2018towards,ozyilmaz2018idmob,hynes2018demonstration}). However, this approach limits the flexibility of data validation and processing on-chain, and performing these operations off-chain introduces additional security threats. For the second challenge, prior works have proposed to use homomorphic encryption~\cite{zheng2018scalable}, functional encryption~\cite{koutsos2021agora}, and secure multi-party computation~\cite{liang2021omnilytics} to protect data privacy. Nonetheless, most existing protocols require a trusted third party to generate and distribute keys, which is often not available in practice; and when further processing of the data is needed (say weighted sum), the transactions are executed on chain in~\cite{koutsos2021agora,liang2021omnilytics}, causing low efficiency and high cost. Other works have proposed to achieve private processing of traded data using differential privacy and trusted execution environments (TEE)~\cite{niu2018unlocking,hynes2018demonstration}. This incurs performance loss of ML models due to inaccurate computation, and potentially suffers from security attacks on TEE.     

To overcome the above shortcomings of the current solutions, we propose \scheme, which to the best of our knowledge is the first smart contract-based data market for trading ML models that simultaneously achieves 1) model privacy for the model owner (\mr); 2) data privacy for the data owner (\dr); 3) robustness against poisonous models from malicious \drs; and 4) efficient and cheap execution with on-chain storage and computation complexities that do not scale with model size. Specifically, to protect model privacy, the \mr publishes its initial model encrypted using multiplicative and additive masks, using which each \dr computes encrypted gradient using its private data. \scheme speeds up smart contract execution through offloading the data storage and processing onto off-chain servers. To protect data privacy, each \dr secret shares its computed gradient with the servers, and a lightweight distributed zero-knowledge (ZK) proof is reconstructed on-chain, verifying the validity of the gradient through some predicate represented as an arithmetic circuit. We adopt the SNIP framework~\cite{corrigan2017prio} to construct the distributed ZK proof, and replace the original additive secret sharing with Shamir secret sharing to counter potentially malicious servers. Next, shares of valid gradients are aggregated at each server, which are fetched and verified against the on-chain commitments by the \mr. Finally, the \mr recovers and decrypts the aggregated gradient and uses it to update the model. We note that in \scheme, only scalar functions (i.e., binary validation results and commitments) of the model and the gradients are stored and computed on-chain, rendering the on-chain cost to be independent of the model size, substantially reducing the execution time and gas cost. 

We implement \scheme as an Ethereum smart contract. Extensive experiments empirically demonstrate the effectiveness of \scheme in training ML models in presence of malicious \drs, and its significant advantages in reducing the on-chain gas cost (only 2\% of the fully on-chain solution, and $3\% \sim 24\%$ of other rollup solutions) and execution time over baseline protocols.

\section{Related Works}
\subsection{Federated learning on blockchains}
Federared learning (FL)~\cite{mcmahan2017communication} is an emerging collaborative learning paradigm, in which a central server, collects from distributed clients local models trained over their private data, and aggregates them into a global model. Recent works have proposed to implement FL over blockchains, especially to counter security threats from malicious server (see, e.g.,~\cite{lu2019blockchain,shayan2020biscotti,zhao2020privacy,li2020blockchain,nguyen2021federated,kim2019blockchained,lyu2020towards}). Blockchain-based FL frameworks can be generally categorized into two types, depending on whether modifications are needed for the underlying consensus layer of blockchain, or FL is implemented as a smart contract application. For the first type, for instance, a committee formation and leader selection mechanism for blockchain consensus is proposed based on the qualities of the locally trained models~\cite{lu2019blockchain}; in~\cite{li2020blockchain,shayan2020biscotti}, a committee of nodes verify the validity of the local models, and submit the aggregation of valid models to include in a block. In contrast, smart contract-based FL systems~\cite{short2021execution,Biscotti,ConsensusFL,PPBFL,BAFFLE} implement all the model verification and aggregation operations as codes in smart contracts, leaving the underlying consensus layer untouched. We adopt the smart contract-based approach to design our data market, such that it can be flexibly utilized with any layer-1 solutions that support smart contracts.


\subsection{Scaling techniques for smart contracts}
Rollups, as the most promising technique to scale the execution of smart contracts, move the verification of transactions off-chain, significantly reducing the execution time and gas cost on layer-1 blockchain~\cite{thibault2022blockchain}. 
There are mainly two types of rollups: optimistic rollups and zero-knowledge (ZK) rollups. In optimistic rollups, off-chain nodes execute transactions and upload transactions and new state roots to the blockchain, with a fraud-proving scheme allowing blockchain clients to challenge rollup results~\cite{kalodner2018arbitrum}.
In ZK rollups, an off-chain operator performs batch processing of the transactions, and generates validity proofs (by e.g., zk-SNARKs~\cite{ben2014succinct}) to be uploaded on chain.
These proofs provide evidence that the state transitions are indeed carried out executing the transactions. For both optimistic and ZK rollups the transaction data are replicated on blockchain miners.

Developed upon ZK rollups, validiums further move the storage of the transactions off layer-1 blockchain,
further improving scalability and reducing gas cost. Major validium systems, such as StarkEx \cite{StarkEx}, ensure off-chain data availability by establishing a data availability committee (DAC), which consists of a trusted group of entities responsible for storing state copies and providing evidence of data availability. 
While the adoption of DAC simplifies the system implementation, the security of the system relies on trust in DAC members. There is a potential risk for the transaction data to become tampered or unavailable, if a certain number of DAC nodes are corrupted. 

\begin{table}[htbp]
	\centering  
	\caption{Comparison of smart contract implementations for ML data market.}  
	\label{table1}  
        \resizebox{1\columnwidth}{!}{
        \begin{tabular}{|c|c|c|c|c|c|c|}
		\hline  
		& & & & & &\\[-6pt]  
		 \textbf{Schemes}&\textbf{Security}&\textbf{\makecell{Data \\ Privacy}}& \textbf{\makecell{Model \\ Privacy}}&\textbf{\makecell{Computation \\ Efficiency}}&\textbf{\makecell{Storage \\ Efficiency}}&\textbf{\makecell{Gas \\ Cost}} \\  
		\hline
		& & & & & &\\[-6pt]  
		On-chain &\checkmark&$\times$&$\times$&$\times$&$\times$&$\times$ \\
		\hline
    & & & & & &\\[-6pt]  
\texttt{OmniLytics}~\cite{liang2021omnilytics}
&\checkmark&\checkmark&\checkmark&$\times$&$\times$&$\times$ \\
		\hline
  & & & & & &\\[-6pt]  
		ZK rollups&\checkmark&$\times$&$\times$&\checkmark&$\times$&$\times$ \\
		\hline
  & & & & & &\\[-6pt]  
		Validiums&$\times$&$\times$&$\times$&\checkmark&\checkmark&\checkmark \\
		\hline
  & & & & & &\\[-6pt]  
		\scheme (Proposed)&\checkmark&\checkmark&\checkmark&\checkmark&\checkmark&\checkmark \\
		\hline
	\end{tabular}}
\end{table}

We can build smart contract-based data market with ZK rollups and validiums as scaling solutions. Table~\ref{table1} compares multiple implementations in terms of security, data and model privacy, computation and storage efficiency, and gas cost.
The proposed \scheme scheme offloads the storage of transaction data onto off-chain servers to further save storage and gas costs, compared with solutions with on-chain storage.  
\scheme utilizes model perturbation and Shamir secret sharing to respectively guarantee model and data privacy. Furthermore, verifiable secret sharing with on-chain commitment and robust secret reconstruction are leveraged in \scheme to maintain security against malicious \drs and off-chain servers.

\section{Problem Description}

\subsection{Data market for training ML models}
We consider the problem of data trading over a data market, for machine learning applications (e.g., prediction and classification). There are two types of players on the market:
\begin{itemize}
\item {\bf Model owner (\mr):} the users who would like to obtain an ML model for certain tasks, but has only a limited amount of local data. \mr intends to purchase more training data through the data market to improve the model performance;
\item {\bf Data owner (\dr):} the users who are willing to contribute the utility of their private data for model training, and in return obtain payoffs from the \mr. 
\end{itemize}

Consider $N$ \drs who contribute to the model training. Each \dr $n$ has locally a dataset ${\cal S}_n =\{(\vct{x}^{(1)}_n,\vct{y}^{(1)}_n),\ldots,(\vct{x}_n^{(M_n)},\vct{y}_n^{(M_n)})\}$ of $M_n$ samples. Each sample $(\vct{x}_n^{(m)},\vct{y}_n^{(m)})$ consists of an input vector $\vct{x}_n^{(m)} \in \mathbb{R}^d$ and its label $\vct{y}_n^{(m)} \in \mathbb{R}^p$ for some input dimension $d$ and output dimension $p$. The goal is to train a model (e.g., a neural network) with parameters $\vct{W}$ to minimize the objective
\begin{align}\label{eq:obj_func}
    F(\vct{W}) =  \frac{1}{M}\sum_{n=1}^N \sum_{m=1}^{M_n}{\cal L}(\vct{W};(\vct{x}_n^{(m)},\vct{y}_n^{(m)})),
\end{align}
for some loss function ${\cal L}$, where $M =\sum_{n=1}^N M_n$.


The optimization is performed using gradient descent over multiple iterations. With the initial model parameters $\vct{W}^{(0)}$, in each iteration $t=1,2,\ldots$, the model is updated as 
\begin{align}\label{eq:GD}
    \vct{W}^{t} = \vct{W}^{t-1} - \eta\vct{G}^{t-1},
\end{align}
where $\vct{G}^{t-1} = \frac{\partial F(\vct{W}^{t-1})}{\partial \vct{W}^{t-1}} = \frac{1}{M}\sum_{n=1}^N \sum_{m=1}^{M_n}\frac{\partial {\cal L}(\vct{W}^{t-1}; (\vct{x}_n^{(m)},\vct{y}_n^{(m)}))}{\partial \vct{W}^{t-1}}$ is the gradient computed from \drs' data, and $\eta$ is the learning rate.


 

One way to implement this training process is for a centralized party (e.g., a cloud server) to collect all the data from the \drs and carry out training. However, various privacy, security, and trust issues may arise. Firstly, uploading private data directly causes the \drs to surrender the exclusive ownership of their data, which is highly undesirable; Second, selfish or malicious \drs can provide garbage or poisoned data, to obtain undeserved rewards or harm the model performance; Finally, the centralized party becomes a single vulnerability of the system, such that once compromised, no data privacy, model performance, and trading legitimacy can be guaranteed. This motivates us to design a data market that \emph{achieves accurate and efficient training of ML models, with minimum privacy leakage and maximum trustworthy guarantees}.




\subsection{Security goals}\label{sec:sec_goals}
We explicitly list the desired privacy and security requirements as follows:

\begin{itemize}[leftmargin=*]
    \item {\bf Avoiding single point of failure.} The functionalities of a data market should be realized by a decentralized implementation, for which corrupting a part of system resource will not reveal private data and  compromise the correctness of the data trading logic. 
    
    \item {\bf Model privacy (for \mr).} An \mr would like to keep its model private from all other parties, before and after the training process.
    
    \item {\bf Data privacy (for \dr).} A \dr would like to maintain the ownership of its local data, and keep them private from all other parties.
    
    \item{\bf Security against malicious \drs.} The data market should verify the legitimacy of the information provided by the \drs, protecting the model from being poisoned by ill-formed results from malicious \drs.  
    
    \item {\bf Security against malicious \mr.} The data market should enforce that honest \drs who faithfully follow the protocol get properly compensated for their contributions to the model training.
\end{itemize}

\subsection{Solution overview}
We propose \scheme, a decentralized data market system for ML applications, through developing a blockchain smart contract \contract.
The benefit of designing \scheme as a smart contract is two-folded: 1) Security: it harnesses the strong security of the underlying decentralized blockchain system in reliably realizing specified trading logics; 2) Flexibility: \scheme can be readily deployed on any blockchain system that supports smart contracts, without any modification to the low-level consensus protocol.


To protect \mr's model privacy and \drs' data privacy, we leverage the \emph{federated learning with model perturbation} framework in~\cite{yang2020computation}, which encrypts the model parameters from \mr, and has each \dr only upload the gradient computed from its private data (instead of the data itself). To defend malicious \drs who may upload arbitrarily invalid results, \scheme checks the validity of uploaded gradients using some validation predicate $\texttt{Valid}(\cdot)$ specified by the \mr.

Considering the large sizes of ML models and corresponding gradients, performing all operations on-chain poses heavy storage and computation workloads on the blockchain verifiers, which leads to slow processing and high transaction fees. \scheme further leverages \emph{off-chain} storage and processing to speed up the contract execution and reduce trading costs. Specifically, each \dr secret shares its gradient with multiple off-chain servers, who collectively run a zero-knowledge verifiable computation protocol in~\cite{corrigan2017prio} to check the validity of the gradient, and aggregate the valid ones. The correctness of the off-chain computations are verified on-chain, using model commitments provided by \drs.

We instantiate \scheme as an Ethereum smart contract \ethcontract with 400 lines of Solidity code. We experimentally demonstrate its superiority in security, efficiency, and affordability.

\section{Preliminaries}\label{sec:SFL}

\subsection{Cryptographic primitives}\label{sec:crypto}
\noindent {\bf Arithmetic circuit.} 
An arithmetic circuit ${\cal C}$ over a finite field $\mathbb{F}_q$ of prime order $q$ is a computational graph whose vertices consist of input, addition, and multiplication gates. The input gates hold the values of the inputs to the circuit ${\cal C}$, and the output wire of an addition or multiplication gate carries the addition or multiplication of the input wires for that gate. 

\noindent {\bf Shamir secret sharing.} A $(T,K)$ threshold Shamir secret sharing is a scheme that creates $K$ shares of a secret $\vct{s} \in \mathbb{F}_q^m$ of length $m$, such that any $T$ shares reveal no information about $\vct{s}$; and any $T+1$ shares can be used to perfectly reconstruct $\vct{s}$. The basic scheme consists of the following algorithms.
\begin{itemize}
    \item $([\vct{s}]_i)_{i=1}^K \leftarrow \texttt{SS.share}(\vct{s},T,K)$: From $\mathbb{F}_q^m$, $T$ noises $\vct{z}_1,\ldots,\vct{z}_T$ are sampled uniformly to construct the polynomial $\vct{u}(x) = \vct{s} + \vct{z}_1 x+\cdots+\vct{z}_T x^T$. Then the $i$th share $[\vct{s}]_i$ is generated as $[\vct{s}]_i = \vct{u}(i)$, for $i=1,\ldots,K$.
    \item $\vct{s} \leftarrow \texttt{SS.recon}(\{[\vct{s}]_i: i \in {\cal R}\})$: For a subset of shares in ${\cal R} \subseteq [K]\triangleq \{1,\ldots,K\}$ with $|{\cal R}| > T $, this algorithm interpolates $\vct{u}(t)$ from the shares $\{[\vct{s}]_i: i \in {\cal R}\}$, and reconstructs the secret $\vct{s} = \vct{u}(0)$.
\end{itemize}

The reconstruction process can be strengthened to correctly recover the secret $\vct{s}$, even when up to $\lfloor \frac{K-T-1}{2} \rfloor$ shares are arbitrarily erroneous. That is
\begin{itemize}
    \item $\vct{s} \leftarrow \texttt{SS.robustRecon}(\{[\vct{s}]_i: i \in [K]\})$: We can regard the $K$ shares collectively as a codeword of a $[K,T+1]$ Reed-Solomon code~\cite{wicker1999reed}, for which error-correction decoding~(e.g., Gao's decoder~\cite{gao2003new}) can be utilized to correct up to $\lfloor \frac{K-T-1}{2} \rfloor$ symbol errors.
\end{itemize}

For verifiable Shamir secret sharing, a check string $\vct{c}_{\vct{s}}$ is generated along with the $K$ shares, i.e., $(([\vct{s}]_i)_{i=1}^K, \vct{c}_{\vct{s}})\leftarrow \texttt{SS.share}(\vct{s},T,K)$. The validity of a share can be verified using $\vct{c}_{\vct{s}}$ through the following \texttt{verify} algorithm. 
\begin{itemize}
    \item $v \in \{0,1\} \leftarrow \texttt{SS.verify}(\vct{v},i,\vct{c}_{\vct{s}})$: This algorithm outputs $1$ if $\vct{v}$ is indeed the $i$th share of $\vct{s}$, i.e., $\vct{v}=[\vct{s}]_i$. For any  $T+1$ valid shares $\vct{v}_1,\ldots,\vct{v}_{T+1}$, we have $\vct{s} = \texttt{SS.recon}(\vct{v}_1,\ldots,\vct{v}_{T+1})$.
\end{itemize}

\noindent {\bf Secret-shared non-interactive proof (SNIP).} A SNIP protocol~\cite{corrigan2017prio} allows a client to prove to a group of servers, that its private input $\vct{s} \in \mathbb{F}_q^m$ is valid, i.e., $\texttt{Valid}(\vct{s})=1$ for some predicate $\texttt{Valid}(\cdot)$, which is represented as an arithmetic circuit ${\cal C}$. The protocol would like to keep the proof zero-knowledge, such that for an honest client, any proper subset of curious servers know nothing about $\vct{s}$ other than that $\texttt{Valid}(\vct{s})=1$; and for a malicious client whose input is invalid, as long as all servers are honest, they will reject the input with overwhelming probability.

As the first step, the client sends an additive secret share $\vct{s}^{(i)}$ to server $i$, such that $\sum_{i} \vct{s}^{(i)} = \vct{s}$. Next, the client evaluates the circuit ${\cal C}$ using $\vct{s}$. Let $M$ be the total number of multiplication gates in ${\cal C}$, and $u_t$, $v_t$ be the values of the input wires of multiplication gate $t \in [M]$. Then, the client interpolates the polynomials $f(x)$ and $g(x)$ such that $f(t)=u_t$ and $g(t)=v_t$ for all $t \in [M]$, and constructs a polynomial $h(x) = f(x) \cdot g(x)$. Client generates additive secret shares of the coefficients of $h(x)$, and shares a polynomial $h^{(i)}(x)$ with server $i$. Using $\vct{s}^{(i)}$ and $h^{(i)}(x)$, server $i$ computes $\{f^{(i)}(t):t \in [M]\}$ and $\{g^{(i)}(t):t \in [M]\}$, where $f^{(i)}(x)$ and $g^{(i)}(x)$ are polynomials whose coefficients are shares of those of $f(x)$ and $g(x)$ respectively. Next, server $i$ interpolates $\hat{f}^{(i)}(x)$ and $\hat{g}^{(i)}(x)$ from the $M$ evaluation points.  

To verify that $h(x) = \hat{f}(x) \cdot \hat{g}(x)$ holds so that the client's input is correctly processed by ${\cal C}$, a polynomial identity test is performed where for a randomly chosen $r$, it is checked whether $\sum_{i} (h^{(i)}(r)-(\hat{f} (r)\cdot \hat{g}(r))^{(i)}) = 0$ holds. Specifically, server $i$ utilizes a Beaver triple~\cite{beaver1992efficient} to obtain the $i$th share of the product $(\hat{f} (r)\cdot \hat{g}(r))^{(i)}$ from the shares $\hat{f}^{(i)}(r)$ and $\hat{g}^{(i)}(r)$ (see Appendix~\ref{sec:beaver} for a description of Beaver triples). Once the identity test is passed, the servers exchange the shares of the output wire to obtain the circuit output. 

\subsection{Federated learning with model perturbation}\label{sec:FL_model}
Federated learning (FL) is an emerging collaborative learning paradigm, where a group of clients (analogous to the \drs in our setting) each with some local private data, collaborate to train an ML model with the coordination of a central server. With the objective of minimizing the function $F(\vct{W})$ in (\ref{eq:obj_func}), each client $n$ optimizes a local objective $F_n(\vct{W}) =\frac{1}{M_n}\sum_{m=1}^{M_n}{\cal L}(\vct{W};(\vct{x}_n^{(m)},\vct{y}_n^{(m)}))$.
Specifically, in iteration $t$, each client $n$ computes the local gradient $\vct{G}_n^{t-1} = \frac{\partial F_n(\vct{W}^{t-1})}{\partial \vct{W}^{t-1}}$ and uploads it to the server, who aggregates received gradients with weight $p_n = \frac{M_n}{\sum_{n=1}^N M_n}$ and updates the model as follows.
\begin{align}\label{eq:GD}
    \vct{W}^{t} = \vct{W}^{t-1} - \eta \sum_{n=1}^N p_n \vct{G}_n^{t-1}.
\end{align}
FL is considered privacy-preserving to some extend, as the raw training data never leave the clients.


For feed-forward neural networks $\vct{W}$ with ReLU activation, and quadratic loss ${\cal L}(\vct{W};(\vct{x},\vct{y})) = \frac{1}{2}||\hat{\vct{y}} - \vct{y}||_2^2$ where $\hat{\vct{y}}$ is the predicted output of $\vct{x}$ using $\vct{W}$, the following model perturbation mechanism was proposed in~\cite{yang2020computation} such that the model parameters are kept private from the clients, and the server can decrypt the aggregated gradient losslessly from the local gradients computed on the perturbed model.

\noindent {\bf Model perturbation.} For a feed-forward neural network of $L$ layers, we denote the model parameters in layer $l \in [L]$ as $\vct{W}^{(l)} \in \mathbb{R}^{n_l \times n_{l-1}}$, where $n_l$ is the number of neurons in layer $l$.
The server encrypts the model parameters $\vct{W}$ into $\widetilde{\vct{W}}$ before distributing them to the clients such that
\begin{equation}\label{eq:model_enc}
\widetilde{\vct{W}}^{(l)} =\begin{cases}
\vct{R}^{(l)}\circ \vct{W}^{(l)}, & 1\leq l\leq L-1,\\
\vct{R}^{(l)}\circ \vct{W}^{(l)}+\vct{R}^{(a)}, & l=L,
\end{cases}
\end{equation}
where $\circ$ denotes the Hadamard product. For each layer $l \in [L]$, the multiplicative masks are generated as
\begin{equation} \label{2}
R_{ij}^{(l)} = \begin{cases}
    r_i^{(1)}, & l=1,\\
    r_i^{(l)}/r_j^{(l-1)}, &2\leq l\leq L-1,\\
    1/r_j^{(L-1)}, &l=L,
\end{cases}
\end{equation}
for some positive noise vector ${\vct{r}}^{(l)}=[r_1^{(l)},\ldots,r_{n_l}^{(l)}]^\top \in \mathbb{R}_{>0}^{n_l}$ randomly sampled at the server.
For the last layer $L$ of $\vct{W}$, for some additive noise vectors ${\vct{r}}^{(a)}=[r_1^{(a)},\ldots,r_{n_L}^{(a)}]^\top \in \mathbb{R}^{n_L}$ and $\vct{\gamma}=[\gamma_1,\ldots,\gamma_{n_L}]^\top \in \mathbb{R}^{n_L}$, both generated randomly at the server, we have $R_{ij}^{(a)} = \gamma_i \cdot r_i^{(a)}$. The server publishes the encrypted model $\widetilde{\vct{W}}=(\widetilde{\vct{W}}^{(l)})_{l=1}^L$ and the noise $\vct{r}^{(a)}$ to the clients.

\noindent {\bf Local gradient computation.} 
In forward propagation, for each sample $(\vct{x},\vct{y}) \in {\cal S}_n$ at client $n$, the output vector of layer $l$, $\widehat{\vct{y}}^{(l)}=[\widehat{y}^{(l)}_1,\ldots,\widehat{y}^{(l)}_{n_l}]^\top \in \mathbb{R}^{n_l}$ is computed as 
\begin{equation}
\widehat{\vct{y}}^{(l)} =\begin{cases}
\vct{x}, &l=0,\\
\textup{ReLU}(\widetilde{\vct{W}}^{(l)}\widehat{\vct{y}}^{(l-1)}), & 1\leq l\leq L-1,\\
\widetilde{\vct{W}}^{(l)}\widehat{\vct{y}}^{(l-1)}, & l=L.
\end{cases}
\end{equation}

In backward prorogation, client $n$ computes for each layer $l \in [L]$:
\begin{align}\label{eq:local_comp}
     \widetilde{\vct{G}}_n^{(l)} &=\nabla F_n(\widetilde{\vct{W}}^{(l)}),\\
    \widetilde{\vct{\sigma}}_n^{(l)}&=\frac{1}{M_n}\sum_{m=1}^{M_n}\vct{\sigma}_{(\vct{x}_n^{(m)},\vct{y}_n^{(m)})}^{(l)},\\
    {\vct{\beta}}_n^{(l)}&=\frac{1}{M_n}\sum_{m=1}^{M_n}{\vct{\beta}}_{(\vct{x}_n^{(m)},\vct{y}_n^{(m)})}^{(l)},
\end{align}
where for $\alpha = \widehat{y}_1^{(L-1)} + \cdots + \widehat{y}_{n_{L-1}}^{(L-1)}$, we have
\begin{align}
{\vct{\sigma}}^{(l)}_{(\vct{x},\vct{y})}&=\alpha\frac{\partial\widehat{\vct{y}}^{(L)}}{\partial\widetilde{\vct{W}}^{(l)}}+\left(\frac{\partial\mathcal{\mathcal{L}}(\widetilde{\vct{W}};(\vct{x},\vct{y}))}{\partial\widehat{\vct{y}}^{(L)}}\right)^\top \frac{\partial\alpha}{\partial\widetilde{\vct{W}}^{(l)}},\\
{\vct{\beta}}^{(l)}_{(\vct{x},\vct{y})}&=\alpha\frac{\partial\alpha}{\partial \widetilde{\vct{W}}^{(l)}}.
\end{align}

Finally, for $\widetilde{\vct{\sigma}}_n^{(l)} = (\widetilde{\vct{\sigma}}_{n,i}^{(l)})_{i=1}^{n_L}$, client $n$ computes $\vct{\sigma}_{n,i}^{(l)} = r_i^{(a)} \widetilde{\vct{\sigma}}_{n,i}^{(l)}$, for all $i \in [n_L]$, and sends $\vct{Q}_n = ( \widetilde{\vct{G}}_n^{(l)},({\vct{\sigma}}_{n,i}^{(l)})_{i=1}^{n_L},{\vct{\beta}}_n^{(l)})_{l=1}^L$ to the server.

\noindent {\bf Gradient decryption.}
Server aggregates the results from the clients to compute
\begin{align*}
 \widetilde{\vct{G}}^{(l)} &= \sum_{n=1}^N \frac{M_n}{M} \widetilde{\vct{G}}^{(l)}_n,\\
 \vct{\sigma}_{i}^{(l)} &= \sum_{n=1}^N \frac{M_n}{M}  \vct{\sigma}_{n,i}^{(l)}, \quad i = 1,\ldots,n_L,\\
 \vct{\beta}^{(l)} &= \sum_{n=1}^N \frac{M_n}{M}  \vct{\beta}_{n}^{(l)}.
\end{align*}

Server uses its private variables $\vct{R}^{(l)}$ and $\vct{\gamma}$ to recover the gradient for layer $l$:
\begin{equation}\label{eq:decrypt}
    \vct{R}^{(l)}\circ \!\big(\widetilde{\vct{G}}^{(l)} \!-\!{\sum\limits_{i=1}^{n_L}}{\gamma_i}\vct{\sigma}_i^{(l)}+(\vct{\gamma} \circ \vct{r}^{(a)})^\top (\vct{\gamma} \circ \vct{r}^{(a)})\vct{\beta}^{(l)}\big) \!\!=\! \nabla F(\vct{W}^{(l)}).
\end{equation}

\subsection{Blockchain and smart contract}

Introduced by Satoshi Nakamoto~\cite{nakamoto2008bitcoin}, blockchain is a specialized form of distributed ledger maintained by Internet-connected nodes. 
A blockchain consists of a sequence of blocks  $(b_1, b_2, \dots, b_n)$, where each block \( b_i \) contains a set of transactions \( T_i \) and a pointer $h_{i-1}$ to its previous (parent) block. Blockchain technology offers security through its immutable ledger design using cryptographic primitives. The blocks are chained together such that the cryptographic hash of the parent block is sealed in the header of the child block. Hence, any modification to a block's content would necessitate recalculations of the hashes for all subsequent blocks, making unauthorized alterations computationally prohibitive.
Beyond its cryptographic foundation, robust consensus mechanisms like Proof-of-Work~\cite{nakamoto2008bitcoin} and Proof-of-Stake~\cite{saleh2021blockchain} were designed for all the nodes in the network to reach an agreement on the ledger records, even when a subset of participants are controlled by adversary. Blockchain's fundamental principles of transparency, immutability, and traceability are invaluable for digital data markets. Through its secure data sequencing, blockchain guarantees data integrity and uniqueness, positioning it as an advanced solution to challenges in conventional data market infrastructures.

Ethereum \cite{ETH} was the first blockchain platform to introduce the notion of ``smart contract'', which is a special type of account on Ethereum blockchain. A smart contract contains a program that defines certain execution rules for state transition, and interacts with other users' accounts through transactions. Smart contract codes are immutable once deployed on-chain, and are executed automatically upon meeting specific criteria. Blockchain verifiers execute the smart contracts, and charge gas fees from the blockchain users according to the number of performed operations.


\section{\scheme Data Market}
\subsection{System overview}
\begin{figure}[htbp]
   \centering
   \includegraphics[width=\linewidth]{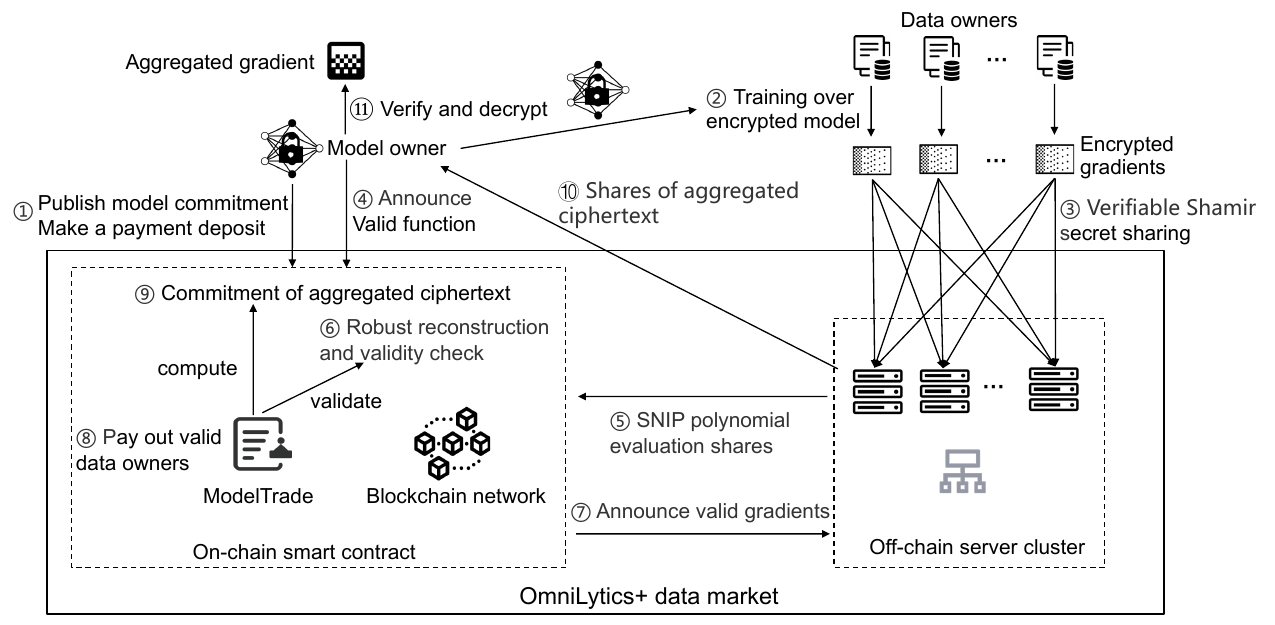} 
   \caption{System overview of the proposed \scheme data market. \scheme consists of an on-chain smart contract \contract and an off-chain server cluster. The interactions between different parties in a data trading session are illustrated and numbered by their orders of occurrence.}
   \label{fig:system}
 \end{figure}

The proposed \scheme data market consists of an on-chain smart contract \contract and a cluster of $K$ off-chain servers. As shown in Figure~\ref{fig:system}, the overall data trading functionality is implemented by \contract, which executes on the verifiers of the underlying blockchain network.
To speed up the contract execution and reduce the transaction fees, \scheme further augments \contract with off-chain servers to offload storage and computation burden from blockchain.
Specifically, using the encrypted model of \mr whose commitment is posted on chain, each \dr computes its local gradient, and secret shares it with the off-chain servers. The servers collectively verify the validity of the gradient from the shares, such that the final (light-weight) validation result is reconstructed on chain. This also triggers the payments to the validated \drs for their contributions to model training. Next, secret shares of all validated gradients are aggregated at each server to form a single share of aggregation, and the commitment to the aggregated gradient is computed on chain.
Finally, the aggregated gradient is reconstructed and decrypted by the \mr. 

To verify the validity of a \dr's gradient without leaking its local data, \scheme employs the SNIP protocol between the \dr and the servers, with the validation result posted on chain. To ensure the integrity of the computation results against malicious servers, \scheme utilizes Shamir secret sharing for the SNIP protocol, and adopts robust reconstruction (i.e., Reed-Solomon decoding) for validity check.

\subsection{Validation predicate}
To prevent selfish or malicious \drs from uploading garbage or faulty results, \scheme performs validity check for each \dr's result $\vct{g}$. Specifically, we employ a validation predicate $\texttt{Valid}(\cdot)$ that outputs $1$ when $\vct{g}$ is tested to be valid, and $0$ otherwise. We consider a class of validation predicates that can be represented as an arithmetic circuit over a finite field $\mathbb{F}_q$.
For instance, for the norm threshold test~\cite{sun2019can}, $\texttt{Valid}(\vct{g}) = \mathbbm{1}[||\vct{g}||_2 \leq \rho]$ and only inputs with norms less than the threshold $\rho$ are considered valid. The parameters (e.g., $\rho$ in the norm threshold test) for the validity test are provided by the \mr, which are computed from its local data. 

\subsection{\contract workflow}
A data trading session on \scheme is accomplished through an invocation of the smart contract \contract, which implements a single iteration of the model update as in (\ref{eq:GD}), with the initial model from the \mr, and the gradients collected over participating \drs. The model perturbation techniques described in Section~\ref{sec:FL_model} are employed to protect the model privacy of the \mr. Each \dr performs the SNIP protocol with off-chain servers to verify the validity of its gradient in a privacy-preserving manner. While the original SNIP does not guarantee soundness in presence of malicious servers, we modify it with Shamir secret sharing and utilize robust reconstruction (i.e., Reed-Solomon decoding) to ensure correctness of verification results when (up to a certain number of) off-chain servers are untrusted.

\begin{figure}[htbp]
   \centering
   \includegraphics[width=\linewidth]{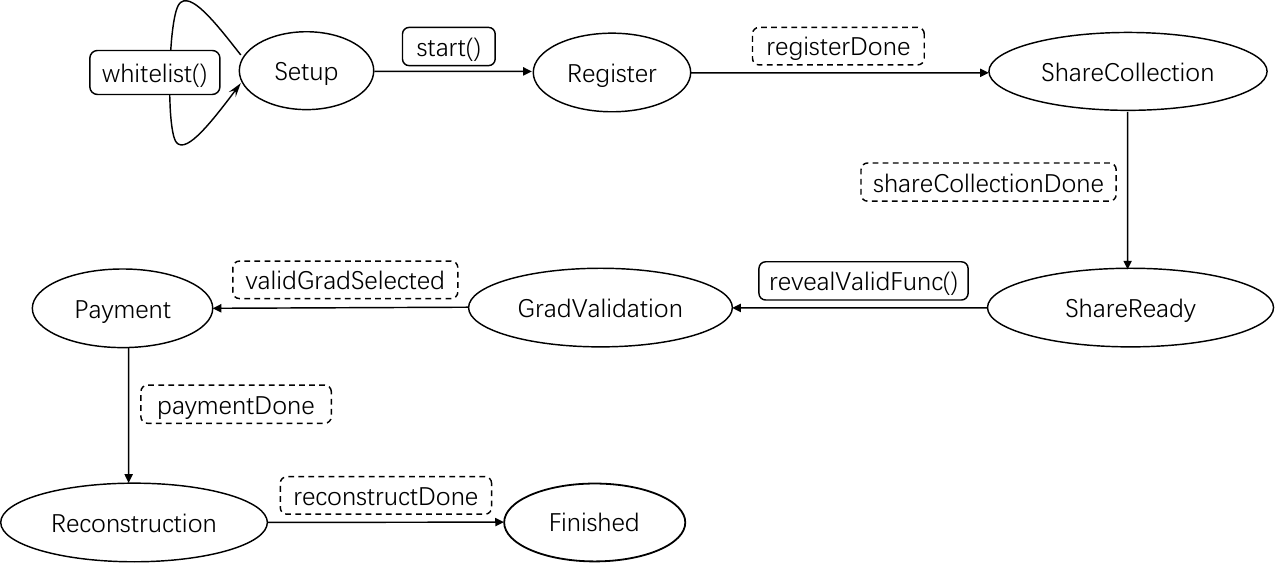} 
   \caption{State transition of the smart contract \contract. The eight states of \contract are represented by ovals. State transitions are triggered by either applying a method (in a solid box), or occurrence of an event (in a dashed box).}
   \label{fig:states}
 \end{figure}

To start, an \mr deploys \contract with training reward deposit on the blockchain. As shown in Figure~\ref{fig:states}, \contract transitions between eight states, i.e., \texttt{Setup}, \texttt{Register}, \texttt{ShareCollection}, \texttt{ShareReady}, \texttt{GradValidation}, \texttt{Payment}, \texttt{Reconstruction}, and \texttt{Finished}. 

\noindent {\bf Step 1: Model publication.}
Upon deployment, \contract is in the \texttt{Setup} state, with a set of \drs the \mr would like to purchase data from specified by a $\texttt{whitelist}()$ method. The \mr encrypts its initial model $\vct{W}$ to obtain an encrypted model $\widetilde{\vct{W}}$ as in (\ref{eq:model_enc}), and hashes $\widetilde{\vct{W}}$ and the generated additive noise vector $\vct{r}^{(a)}$ using Merkle tree into a Merkle root $R(\widetilde{\vct{W}},\vct{r}^{(a)})$. The \mr issues a transaction using the $\texttt{start}()$ method to announce $R(\widetilde{\vct{W}},\vct{r}^{(a)})$, together with the following public parameters:
\begin{itemize}[leftmargin=*]
    \item Required number of data points for each participating \dr to compute its gradient, denoted by $M_0$;
    \item Maximum number of distinct \drs to collect gradients from, denoted by $N$.
\end{itemize}

Executing $\texttt{start}()$ moves \contract into the \texttt{Register} state, and the contract starts to register for the \drs who intend to participate in data trading. The registration is done when $N$ \drs have registered, or a timeout is triggered. 

\noindent {\bf Step 2: Verifiable gradient sharing.} Denote the index set of the registered \drs as ${\cal N}$ with $|{\cal N}| \leq N$. Each \dr $n \in \mathcal{N}$ retrieves the encrypted model $(\widetilde{\vct{W}},\vct{r}^{(a)})$ from the \mr, and verifies its integrity with the Merkle root commitment on chain. Next, the \dr performs back propagation on $\widetilde{\vct{W}}$ and its local data to obtain the encrypted gradient $\vct{Q}_n = (\widetilde{\vct{G}}_n^{(l)},({\vct{\sigma}}_{n,i}^{(l)})_{i=1}^{n_L},{\vct{\beta}}_n^{(l)})_{l=1}^L \in \mathbb{R}^m$ as described in Section~\ref{sec:FL_model}, where the dimension $m = (n_L + 2)w$, and $w=\sum_{l=1}^L n_ln_{l-1}$ is the total number of parameters in the model $\vct{W}$. After proper scaling and quantization, each element of $\vct{Q}_n$ is embedded onto a finite field $\mathbb{F}_q$ with a sufficiently large prime $q$. Next, \dr $n$ performs a $(T,K)$ verifiable Shamir secret sharing to generate $K$ shares $([\vct{Q}_n]_i)_{i=1}^K$ and a commitment $\vct{C}_{n}$ to $\vct{Q}_n$, as $(([\vct{Q}_n]_i)_{i=1}^K, \vct{C}_n)\leftarrow \texttt{SS.share}(\vct{Q}_n,T,K)$. \dr $n$ posts the commitment $\vct{C}_n$ on the \contract contract, and sends the share $[\vct{Q}_n]_i$ to off-chain server $i$, for all $i \in [K]$.

\noindent {\bf Minimizing commitment size.} As the dimension $m$ of $\vct{Q}_n$ is often large due to the large number of parameters in an ML model, utilizing conventional verifiable secret sharing schemes like Feldman's scheme~\cite{feldman1987practical} could incur high computational and monetary costs on the contract. Here we adopt a vector commitment scheme from \cite{jahani2023byzantine} to construct a \emph{constant-size} (does not scale with $m$) commitment $\vct{C}_n$ to $\vct{Q}_n$. Specifically, for a cyclic group $\mathbb{G}$ with a generator $g$, \contract randomly samples an element $\alpha \in \mathbb{F}_q$ and publishes $(g^{\alpha^0},g^{\alpha^1},\ldots,g^{\alpha^{m-1}})$. Each \dr $n$, with the local masks $\vct{Z}_n^{(1)},\ldots,\vct{Z}_n^{(T)}$ to generate its Shamir secret shares, computes and submits to the contract its commitment $\vct{C}_n$ that contains $T+1$ elements in $\mathbb{G}$ as follows:
\begin{align}\label{eq:commit}
    C_n^{(j)} = \begin{cases}
        \prod_{k=1}^m (g^{\alpha^{k-1}})^{\vct{Q}_n[k]}, & j=0, \\
         \prod_{k=1}^m (g^{\alpha^{k-1}})^{\vct{Z}^{(j)}_n[k]}, & j=1,\ldots,T.
    \end{cases}
\end{align}
Each server $i$ verifies its share $[\vct{Q}_n]_i$ with the parameters $(g^{\alpha^0},g^{\alpha^1},\ldots,g^{\alpha^{m-1}})$ and the commitment $(C_n^{(0)},C_n^{(1)},\ldots,C_n^{(T)})$ stored on chain. It checks 
\begin{align}
\prod_{j=0}^{T} (C_n^{(j)})^{i^j} \overset{?}{=} \prod_{k=1}^m (g^{\alpha^{k-1}})^{[\vct{Q}_n]_i[k]}.
\end{align}
If the above check does not pass, server $i$ issues a complaint, and \dr $n$ sends the share $[\vct{Q}_n]_i$ to \contract for further verification.

The \contract contract maintains a set $\mathcal{V}$ of valid \drs. All \drs who have its shares verified by all servers (or the contract in the case where malicious servers file fraudulent complaints on correct shares) are added to $\mathcal{V}$ by the end of this step. It is also important to note that thanks to verifiable secret sharing, it is guaranteed that each honest server has a \emph{correct} share of $\vct{Q}_n$, for all $n \in \mathcal{V}$.

\noindent {\bf Step 3: Private gradient validation.} At the beginning of this step, the \contract contract reveals the arithmetic circuit for evaluating the $\texttt{Valid}(\cdot)$ predicate provided by the \mr. We design this to be done after the \drs' encrypted gradients are shared with off-chain servers, to eliminate the possibility for a malicious \dr to meticulously craft its data to escape the validity check (e.g., performing projected gradient descent to bypass norm-based defenses for injecting model backdoors~\cite{wang2020attack}). 

For the announced $\texttt{Valid}(\cdot)$ circuit, each \dr $n$ executes the SNIP protocol with the $K$ servers, using Shamir secret shares instead of additive shares in the original proposal~\cite{corrigan2017prio}. Specifically, \dr $n$ evaluates $\texttt{Valid}(\vct{Q}_n)$, and obtains a polynomial $h(x)$ of degree at most $2(M-1)$ as described in Section~\ref{sec:crypto}, where $M$ is the number of multiplication gates in the circuit. \dr $n$ applies Shamir secret sharing to the coefficients of $h(x)$, and sends the resulting share polynomial $[h]_i(x)$ to each server $i$ off chain. For each multiplication gate, the data owner also generates a Beaver triple $(a,b,c) \in \mathbb{F}_q$ where $c = a \cdot b$, and sends a triple of Shamir shares $([a]_i, [b]_i, [c]_i)$ to each server $i$. Next, for the identity test, \contract samples and publishes $r \overset{\textup{R}}{\leftarrow} \mathbb{F}_q$. Each server $i$ uploads the share $[h(r)-\hat{f} (r)\cdot \hat{g}(r)]_i$ to \contract, who reconstructs $h(r)-\hat{f} (r)\cdot \hat{g}(r) \leftarrow \texttt{SS.robustRecon}(\{[h(r)-\hat{f} (r)\cdot \hat{g}(r)]_i: i \in [K]\})$ through robust reconstruction. Once the identity test passes, i.e., $h(r)-\hat{f} (r)\cdot \hat{g}(r)=0$, the shares of the output wire of $\texttt{Valid}(\cdot)$ are collected by \contract, upon which robust reconstruction is performed to recover $\texttt{Valid}(\vct{Q}_n)$. \drs with an invalid gradient (i.e., $\texttt{Valid}(\vct{Q}_n)=0$) are removed from the validity set ${\cal V}$. 

At this point, the contract evenly distributes the training reward deposited by the \mr into accounts of the \drs in the validity set ${\cal V}$.

\noindent {\bf Step 4: Aggregated gradient reconstruction.} Based on the validity set ${\cal V}$, each server $i$ aggregates its shares of the valid gradients to obtain a share of aggregated gradient, denoted as $[\vct{Q}_{\Sigma}]_i = \sum_{n \in {\cal V}}[\vct{Q}_n]_i$. The \contract contract computes the commitment $\vct{C}_{{\cal V}}$ to the aggregation of valid gradients, from their individual commitments as follows 
\begin{align}\label{eq:commit_agg}
  C_{{\cal V}}^{(j)} = \prod_{n \in {\cal V}} C_n^{(j)} = \begin{cases}
        \prod_{k=1}^m (g^{\alpha^{k-1}})^{\sum_{n \in {\cal V}}\vct{Q}_n[k]}, & j=0, \\
         \prod_{k=1}^m (g^{\alpha^{k-1}})^{\sum_{n \in {\cal V}}\vct{Z}^{(j)}_n[k]}, & j=1,\ldots,T.
    \end{cases}
\end{align}

For each $i \in [K]$, the \mr retrieves the aggregated gradient share $[\vct{Q}_{\Sigma}]_i$ from server $i$, and verifies its validity against the commitment $\vct{C}_{{\cal V}}$ computed above, via the following check
\begin{align}
\prod_{j=0}^{T} (C_{\cal V}^{(j)})^{i^j} \overset{?}{=} \prod_{k=1}^m (g^{\alpha^{k-1}})^{[\vct{Q}_{\Sigma}]_i[k]}.
\end{align}
Using the shares that pass the above check (at least $T+1$ of them), the \mr reconstructs the gradient aggregation results $\vct{Q}_{\Sigma}$, and computes their averages $\frac{\vct{Q}_{\Sigma}}{|{\cal V}|} =(\frac{1}{|{\cal V}|}\sum_{n \in {\cal V}}\widetilde{\vct{G}}_n^{(l)},(\frac{1}{|{\cal V}|}\sum_{n \in {\cal V}}{\vct{\sigma}}_{n,i}^{(l)})_{i=1}^{n_L},\frac{1}{|{\cal V}|}\sum_{n \in {\cal V}}{\vct{\beta}}_n^{(l)})_{l=1}^L= (\widetilde{\vct{G}}^{(l)},({\vct{\sigma}}_{i}^{(l)})_{i=1}^{n_L},{\vct{\beta}}^{(l)})_{l=1}^L$.

After the execution of the \contract contract, the \mr uses its private variables $\vct{R}^{(l)}$ and $\vct{\gamma}$ to decrypt the plaintext gradient $\nabla F(\vct{W}^{(l)})$ from $\frac{\vct{Q}_{\Sigma}}{|{\cal V}|}$ according to~(\ref{eq:decrypt}), for each layer $l\in [L]$, and updates its ML model with the decrypted gradient.

\section{Analysis}
\subsection{Security analysis}
We analyse the proposed \scheme data market to illustrate how it satisfies the security goals listed in Section~\ref{sec:sec_goals}.

\noindent {\bf Avoiding single point of failure.} Building \scheme on top of a decentralized blockchain system and realizing the data trading logic as a smart contract effectively alleviate the security risk due to single point of failure. Harassing the security guarantees of the underlying blockchain system, the market remains to operate correctly even if a subset of blockchain nodes are compromised. For instance, an implementation of \scheme on top of \ethe would execute the trading logic correctly, even if up to $50\%$ of the blockchain nodes are controlled by adversaries.

\noindent {\bf Model privacy.} The privacy of the model parameters is protected by the model perturbation technique described in Section~\ref{sec:FL_model}. Only the encrypted  model parameters are revealed on the contract, and all the subsequent operations performed on the \drs, the off-chain servers, and the contract are over the ciphertext. As shown in~\cite[Theorem 4]{yang2020computation}, the probability for any party in the \scheme system to obtain true values of model parameters, other than the \mr who has the private key $((\vct{R}_{\ell})_{\ell=1}^L,\vct{\gamma})$, equals to $0$. 

\noindent {\bf Data privacy.} Each \dr $n$ computes a local gradient $\vct{Q}_n$ from its local private data and the \mr's encrypted model. The commitment $C_n^{(0)}$ on $\vct{Q}_n$ is generated as in (\ref{eq:commit}) and revealed in public. The privacy of $\vct{Q}_n$ (and hence the data of \dr $n$) against $C_n^{(0)}$ relies on the Discrete Logarithm (DL) assumption, in the sense that any efficient algorithm in breaking the hiding property of the commitment can be used to solve the DL problem efficiently (see proof of Theorem 1 in~\cite{jahani2023byzantine}). In addition, $\vct{Q}_n$ is also secret shared among, and jointly verified by the off-chain servers. The privacy of $\vct{Q}_n$ is protected by the security guarantees of Shamir secret sharing and the adopted SNIP protocol. Specifically, $\vct{Q}_n$ is perfectly secure against no more than $T$ colluding servers, for all $n \in [N]$. Finally, the aggregated gradient ciphertext $\vct{Q}_{\Sigma} = \sum_{n \in {\cal V}}\vct{Q}_n$ and plaintext $\nabla F(\vct{W}^{(l)}) = \frac{1}{|{\cal V}|}\sum_{n \in {\cal V}}\nabla F_n(\vct{W}^{(l)})$ are respectively recovered on chain and at the \mr, which do not reveal the individual gradients computed from each \dr.

\noindent {\bf Security against malicious \drs.} The validity of the submitted result from each \dr is verified by the $\texttt{Valid}(\cdot)$ predicate. The soundness property of the adopted SNIP protocol guarantees that any malicious \dr who does not submit a valid gradient (according to $\texttt{Valid}(\cdot)$) will be rejected with overwhelming probability, in the case where all servers are honest. 

\noindent {\bf Security against $\boldsymbol{\lfloor \frac{K-T-1}{2} \rfloor}$ malicious servers.} While the original SNIP protocol~\cite{corrigan2017prio} does not promise soundness against malicious servers, the proposed \scheme system enables robustness against malicious servers, via modifying SNIP with Shamir verifiable secret sharing and Reed-Solomon decoding. Specifically, consider up to $A$ servers are malicious and can arbitrarily deviate from the prescribed protocol. By the end of the verifiable gradient sharing step, due to the correctness property of the employed commitment scheme (see \cite[Theorem 1]{jahani2023byzantine}), each honest server is guaranteed to hold a correct share of the gradient $\vct{Q}_n$, for each \dr $n$. This effectively guarantees that in the following SNIP execution for privacy-preserving validity check, at most $A$ of the results submitted from the servers to the contract are erroneous. As both operations conduced on the contract (evaluating the identity test output and circuit output) are Shamir share reconstruction (or polynomial interpolation), Reed-Solomon decoding can be directly utilized to exactly recover the final results as long as $A \leq \lfloor \frac{K-T-1}{2} \rfloor$. 

Moreover, for $A \leq \lfloor \frac{K-T-1}{2} \rfloor$, the number of correct shares of the aggregated gradient $\vct{Q}_{\Sigma}$ collected at the \mr in the last step is at least $K-A \geq \frac{K+T+1}{2} \geq T+1$, which is sufficient for the \mr to correctly reconstruct $\vct{Q}_{\Sigma}$.

\noindent {\bf Security against malicious \mr.} Utilizing the \contract contract, the distribution of training reward to valid \drs is triggered in the end of the private gradient validation step, after the updating of the validity set ${\cal V}$, and is automatically executed by the contract. This effectively eliminates the possibility that a malicious \dr escapes the payment to an honest \dr.
 
\subsection{Efficiency analysis}
As the computational latency and monetary expenses of running \scheme is dominated by the running time and the gas cost of executing the \contract contract, we focus on analyzing the on-chain storage and computation costs of \contract. 

\noindent {\bf Storage cost.} In Step 1 of \contract execution, only a single Merkle root of the encrypted model is stored on chain; In Step 2, to perform verifiable Shamir secret sharing, the contract stores a length-$(T+1)$ commitment for each \dr, incurring a storage cost of $O(NT)$.\footnote{The contract also needs to store $m=(n_L+2)w$ public parameters. However, these parameters can be reused across many contract invocations, and the storage cost per contract call becomes negligible as the number of trading sessions increases.}
In Step 3, the contract reconstructs the validity result for the gradient of each \dr, with a total storage cost of $O(N)$; Finally in the last step, the contract stores the commitment of the aggregated valid gradient, which takes up $O(T)$ storage space. To sum up, the overall storage cost of the developed \contract contract is $O(NT)$.

\noindent {\bf Computational complexity.} In Step 3, the contract performs Reed-Solomon decoding to recover the validity result for the gradient of each \dr, from the results submitted by $K$ servers. This requires a total of $O(N K \log^2 K)$ operations in $\mathbb{F}_q$~\cite{von2013modern}. In the last step, the contract computes the commitment $(C_{{\cal V}}^{(0)},C_{{\cal V}}^{(1)},\ldots,C_{{\cal V}}^{(T)})$ to the aggregated gradient from the individual commitments as in (\ref{eq:commit_agg}), with a complexity of $O(NT)$. Therefore, the total computational complexity of running the \contract contract is $O(N K \log^2K)$.


\section{Evaluation}

\subsection{Implementation}

We implement our proposed \scheme as an Ethereum smart contract \ethcontract, with 400 lines of Solidity code~\cite{Omni}. We deploy the contract on our local Geth Ethereum Testnet~\cite{geth} by Remix IDE~\cite{Remix} and ganache~\cite{ganache}. We implement the \drs and off-chain servers using Go and C++ (for FFT-based polynomial operations, built on the NTL library\cite{libntl}), enabling multiple \drs and servers to operate on a local machine. We quantize the gradient values onto the finite field \( \mathbb{F}_{2^{127}-1} \). For our implementation of ZK rollups, we utilize the zokrates library~\cite{zokrates} to produce Groth'16 zero-knowledge proofs, preserving the transaction data. The rollup operator aggregates these transactions, and creates a ZK-SNARK proof for validation, and then submits this proof alongside a condensed version of the transactions to the Ethereum blockchain. We implement the model training on the \mr and the \drs using Python and Pytorch~\cite{pytorch}. We utilize Web3py library~\cite{Web3py} to connect the off-chain parts and the smart contracts. 


\noindent \textbf{Gradient validation circuit.} We adopt the norm threshold test as the validation method for gradients, i.e., $\texttt{Valid}(\vct{g}) = \mathbbm{1}[||\vct{g}||_2 \leq \rho]$, for some threshold $\rho$ provided by the \mr.
We implement the validation check of the form $||\vct{g}||_2 \leq \rho$ as an arithmetic circuit computing $\left(||\vct{g}||_2-1\right) \times \ldots \times \left(||\vct{g}||_2-\rho\right)$. Specifically, for validation a gradient of size $m$, our circuit outputs $\left( \sum_{i=1}^{m} g_i^2 - 1 \right) \times (\sum_{i=1}^{m} g_i^2-2) \times \ldots \times (\sum_{i=1}^{m} g_i^2-\rho)$. If the value equals zero, this indicates that the the norm of $\vct{g}$ is bounded by $\rho$.


\subsection{Experiments}
\noindent {\bf Robust machine learning.} We run ML experiments on a machine with AMD EPYC 7542 32-Core Processor, NVIDIA GeForce RTX 4090 and 125GB of Memory. We evaluate our scheme on three datasets, which contain both regression and classification tasks, and corresponding network models, as shown in Table 2. We fix the number of data owners in the data market $N = 4$.

\begin{table}[htbp]
	\centering
        \caption{Datasets and corresponding models.}  
	\label{table1}
        \begin{tabular}{|c|c|c|c|}
          \hline
          & & &\\[-6pt]
          \textbf{No.}&\textbf{Dataset}&\textbf{Model}&\textbf{Model Size}\\
          \hline
          & & &\\[-6pt]
          1 & UBMD \cite{moro2014data} & MLP & 7450\\
          \hline
          & & &\\[-6pt]
          2 & MNIST \cite{lecun1998gradient} & LeNet-5 & 60570\\
          \hline
          & & &\\[-6pt]
          3 & CIFAR 10 \cite{krizhevsky2009learning} & ResNet20 & 270338\\
          \hline
        \end{tabular} 
\end{table}

As shown in Figure 3 and Figure 4, we use the Mean Squared Error (MSE) to evaluate performance of the MLP model on the regression task of UBMD dataset, and the test accuracy to evaluate the  ResNet20 model for CIFAR 10 dataset, under 3 scenarios: 1) without malicious \drs; 2) with 25\% of \drs being malicious; and 3) with validation functions against malicious \drs. 
We assume that malicious \drs will upload random noises with the same size as the correct data. 
The experiment results show that for both MLP and ResNet20, the norm threshold test effectively detects the malicious \dr, achieving a comparable performance with the training process without attacks.

\begin{figure}[htbp]
  \centering 
  \begin{minipage}[b]{0.8\columnwidth}
    \includegraphics[width=\linewidth]{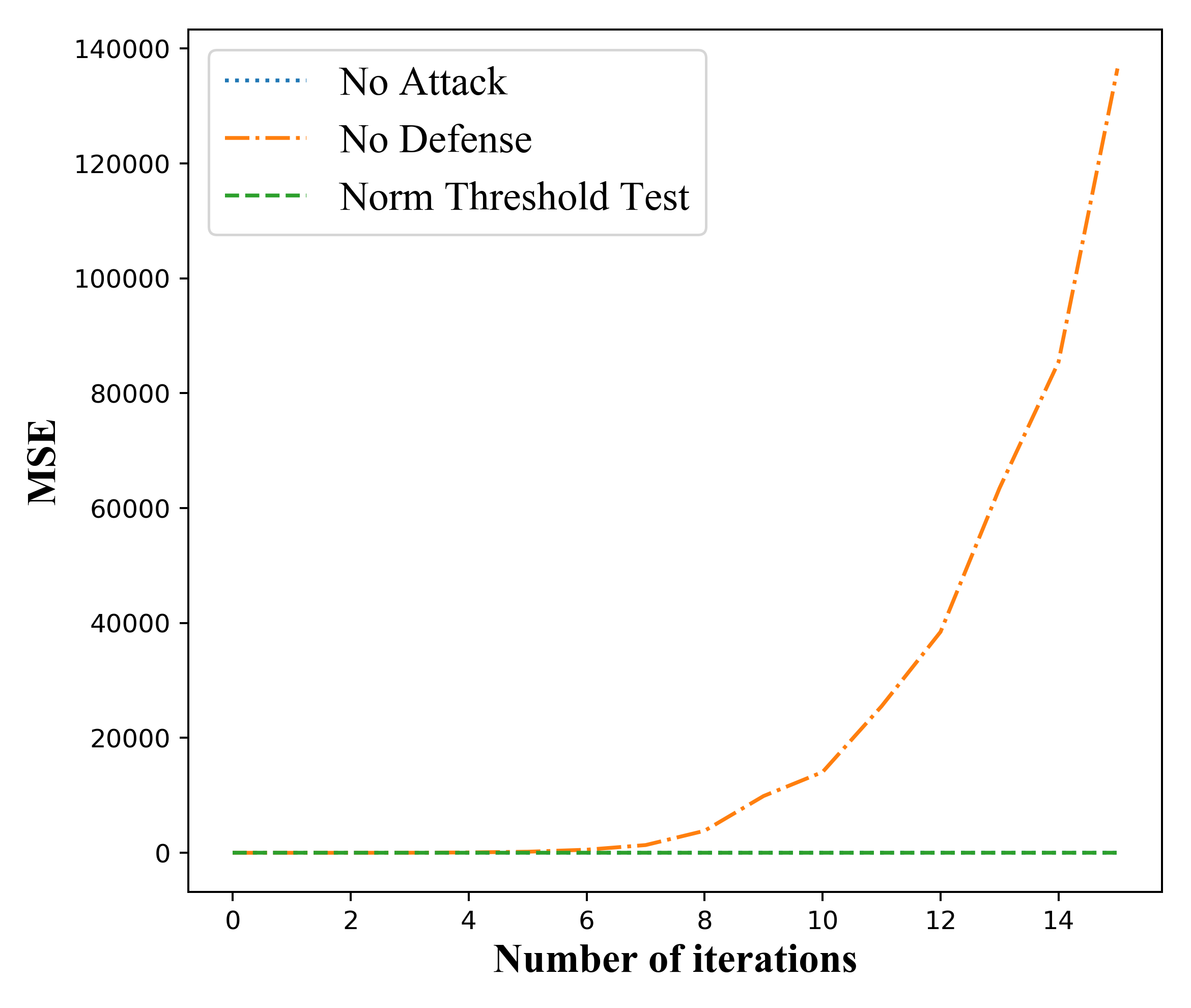}
    \caption{MSE of raining MLP on UBMD.}
    \label{fig:image1}
  \end{minipage}
  \hfill 
  \begin{minipage}[b]{0.8\columnwidth}
    \includegraphics[width=\linewidth]{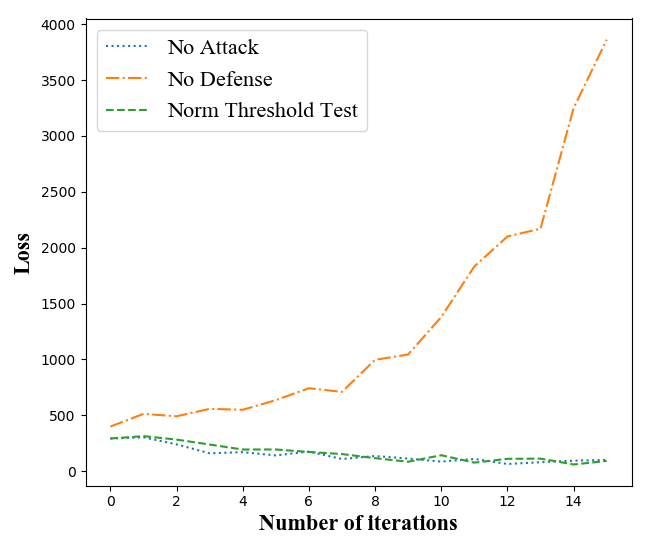}
    \caption{Loss of training ResNet20 on CIFAR 10.}
    \label{fig:image2}
  \end{minipage}
\end{figure}

\noindent {\bf On-chain cost comparison.} We conduct smart contract experiments on a machine with AMD Ryzen R7-5800H CPU @3.20 GHz, 16 GB of memory and 1 TB hard drive and a VMvare workstation with 6.5 GB of memory and 95 GB of hard drive. We initially measure the cost for executing transactions in gas, next, we examine the impact of data owners, servers and model size on the gas needed for the secure data market. After that, we analyze the computation time of the off-chain parts of Omnilytics+.

\noindent \textbf{Omnilytics+ smart contract} 
In this section, we explore the functionalities offered by smart contracts, with a focus on their associated costs. Our \ethcontract smart contract primarily encompasses five functions: `setMerkleRoot`, `storeCommitment`, `storeShares`, `recoverSecret` and `aggregateCommitment`. The `setMerkleRoot` function accepts the MO's Merkle tree root, as specified in Step 1, and saves it as a `bytes32` value. The `storeCommitment` function handles the vector commitments generated by DOs, while `storeShares` processes the identity test results and output wires from Step 3. Additionally, `recoverSecret` facilitates the recovery of the identity test result and output wire within the Solidity smart contract.

We tested different parts of contract methods and the on-chain cost of the whole process on a MLP task with 4 data owners and 5 servers. The results of these tests are summarized in table~\ref{tab:gas_consumption}.

\begin{table}[h]
\centering
\begin{tabular}{l|r}
\toprule
\textbf{Function} & \textbf{Gas Consumption} \\
\midrule
whitelist & 136051 \\
setMerkleRoot & 44096 \\
storeCommitment & 17666388 \\
storeShares & 18685035 \\
aggregateCommitment & 529347 \\
recoverSecret & 2139885 \\
\midrule
\textbf{Total} & \textbf{39200802} \\
\bottomrule
\end{tabular}
\caption{Gas Consumption of Smart Contract Functions}
\label{tab:gas_consumption}
\end{table}

The functions \textit{storeCommitment} and \textit{storeShares} consume the most gas due to the large on-chain storage associated with these operations.

\begin{figure}[htbp]
  \centering 
  \begin{minipage}[b]{0.8\columnwidth}
    \includegraphics[width=\linewidth]{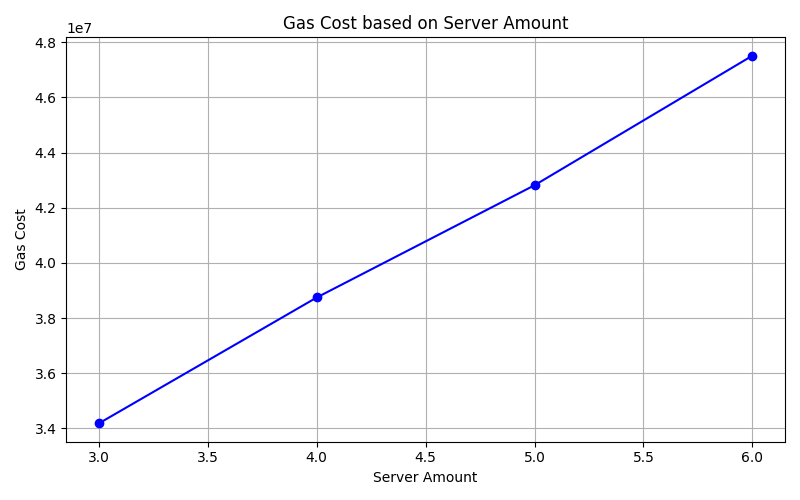}
    \caption{Gas cost for multiple servers.}
    \label{fig:servers_cost}
  \end{minipage}
  \hfill 
  \begin{minipage}[b]{0.8\columnwidth}
    \includegraphics[width=\linewidth]{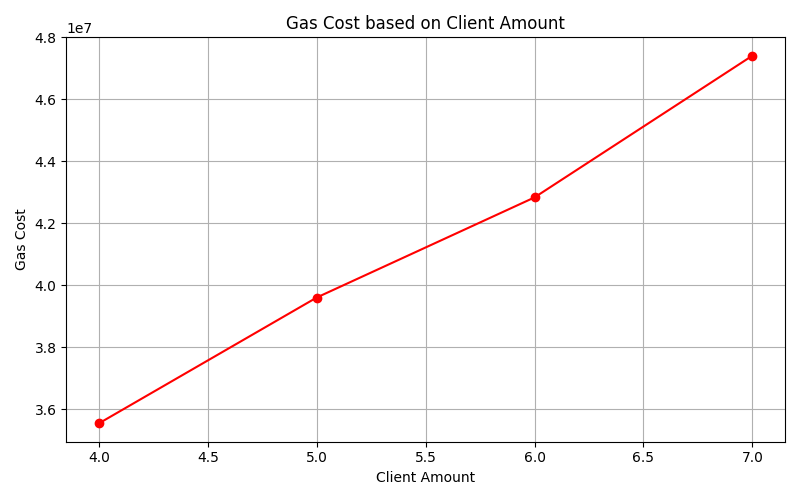}
    \caption{Gas cost for multiple clients.}
    \label{fig:clients_cost}
  \end{minipage}
\end{figure}

In Figure~\ref{fig:servers_cost} and Figure~\ref{fig:clients_cost}, we evaluated on-chain cost with different numbers of servers and clients.

\noindent \textbf{Comparison with different schemes}

\begin{figure}[htbp]
\centering
\includegraphics[width=0.45\textwidth]{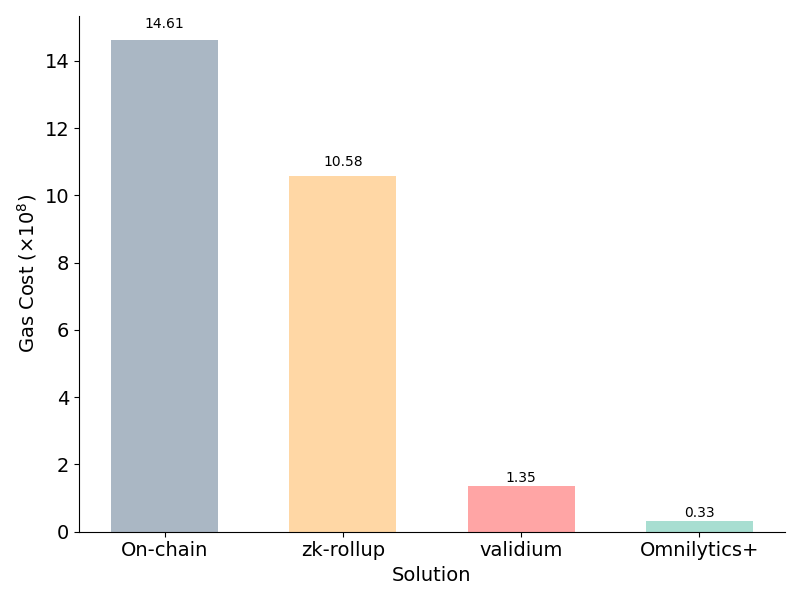}
\caption{Gas cost comparison for different solutions.}
\label{fig:gas_cost_comparison}
\end{figure}

This section presents a comparative analysis of MLP with respect to different total gas costs. In our study, we have compared the gas costs of four distinct clients or solutions. Our findings are summarized in Figure~\ref{fig:gas_cost_comparison}. As observed from our analysis, the On-chain solution demands the highest gas cost, making it the least efficient among the compared solutions. In contrast, Omnilytics+ requires the least gas, indicating its efficiency in conserving resources. The ZK-rollup and Validium solutions fall in between, with Validium being considerably more efficient than ZK-Rollup due to it's off-chain storage.

We achieved gradient uploading, gradient validation and gradient additon in fully on-chain scheme. With the gradient storage and validation procedue fully on-chain, the whole on-chain cost is about 1,460,798,970, which is not practical to achieve data markets on blockchains. In this implemented zk-rollup scheme, instead of computing validation function on-chain, we put the computation of validation process off-chain, and generates ZK-SNARK proofs to verify the validation results. By doing this, the gas cost of the data market is reduced by about 27.6\%. In validium, except from using ZK-SNARK proofs for validating gradients, we also move the storage of gradients off-chain and only store a state proof on smart contract. Because the biggest cost of smart contracts is the storage on them, validium could reduce about 91\% of total on-chain cost. Compared to this schemes, our off-chain scheme is only 3\% of total on-chain cost, meaning most of the gas cost of on-chain data market could be reduced using this way.

\noindent \textbf{Off-chain execution} 
In our experiments concerning the \textit{off-chain} execution, we evaluated the SNIP validation computation technique's performance. We deployed 5 servers and applied gradient validation to each layer of the neural network, which has an approximate size of 2400 units. The average processing time taken across a server is amounted to roughly 43 minutes and 18 seconds. It is deemed acceptable within the context of off-chain computations and making AI model trading possible in the real world.

\section{Conclusion}
We propose \scheme, a smart-contract based decentralized data market for ML applications. \scheme simultaneously achieves model privacy for the model owner, data privacy for the data owners, and robustness against malicious data owners during the trading process. We further utilize off-chain processing to reduce the storage and computation costs of \scheme, reducing the on-chain gas cost and expediting the trading session.

\bibliographystyle{plain}
\bibliography{reference}

\appendix
\section{Beaver Triple}\label{sec:beaver}
In secure multi-party computation of an arithmetic circuit over a finite field $\mathbb{F}_q$, a Beaver triple~\cite{beaver1992efficient} is utilized to efficiently compute the secret share of a multiplication. Specifically, a Beaver triple is a triple of secret-shared values $([a], [b], [c])$, where $a$ and $b$ are randomly chosen from $\mathbb{F}_q$ and $c = a \cdot b$. In evaluating a multiplication gate with input values $\alpha$ and $\beta$, parties holding the input shares $[\alpha]$ and $[\beta]$ compute the output shares $[\alpha \cdot \beta]$ without interaction, using the pre-computed Beaver triple $([a], [b], [c])$. The computation proceeds in the following steps:
\begin{enumerate}
    \item Parties locally compute $[d] = [\alpha-a] = [\alpha] - [a]$, and $[e] = [\beta-b] = [\beta] - [b]$;
    \item All parties open $d=\alpha-a$ and $e=\beta-b$. Note that $d$ and $e$ does not leak any information about $\alpha$ and $\beta$, as $a$ and $b$ respecticaly acts as a one-time pad;
    \item Based on the following equality
    \begin{align*}
        \alpha \cdot \beta &= (d+a) \cdot (e+b)\\
        &=de + db + ae + ab \\ 
        &=de + db + ae + c,
    \end{align*}
    Parties compute sharing of multiplication $[\alpha \cdot \beta] = de + d[b] + e[a] + [c]$.
\end{enumerate}

The above technique of Beaver triple can be used to securely evaluate an arithmetic circuit, for any secret sharing mechanism that satisfies the following requirements~\cite{evans2018pragmatic}:
\begin{itemize}
    \item \emph{Additive homomorphism:} Given the shares $[x]$ and $[y]$, and a public constant $s$, parties can compute the shares $[x+y]$, $[x+s]$, and $[sx]$ locally without interaction;
    \item \emph{Opening:} The sharing $[x]$ can be used to open $x$ to all parties;
    \item \emph{Privacy:} Any adversary cannot obtain any information about $x$ from a share of $[x]$;
    \item \emph{Random input gadgets:} For each input wire of a party $P_i$, it has a random mask $r$, and all parties have a sharing $[r]$. Party $i$ reveals $\delta = x-r$ to all parties (does not leak $x$), and the parties obtain $[x] = [r+\delta]$ from $[r]$ and $\delta$ using the additive homomorphism property.
\end{itemize}

It is easy to verify that the additive secret sharing (used in the original SNIP protocol) and the Shamir secret sharing (used in the proposed \scheme data market) both satisfy the above requirements.

\end{document}